\documentclass{article}

\usepackage[preprint,nonatbib]{neurips_2021}

\usepackage[utf8]{inputenc} % allow utf-8 input
\usepackage[T1]{fontenc}    % use 8-bit T1 fonts
\usepackage{hyperref}       % hyperlinks
\usepackage{url}            % simple URL typesetting
\usepackage{booktabs}       % professional-quality tables
\usepackage{amsfonts}       % blackboard math symbols
\usepackage{nicefrac}       % compact symbols for 1/2, etc.
\usepackage{microtype}      % microtypography
\usepackage{xcolor}         % colors

\usepackage{graphicx}
\usepackage{amsmath}
\usepackage{wrapfig}
\usepackage{biblatex}

\usepackage[caption=false]{subfig}

\newcommand{\norm}[1]{\left\lVert#1\right\rVert}
\newcommand{\pos}[1]{p_{#1}}
\newcommand{\ourmethod}{Picasso}
\newcommand{\hush}[1]{}
\title{\ourmethod{}: Model-free Feature Visualization}

\author{%
  Binh Vu \\
  University of Southern California\\
  \texttt{binhlvu@usc.edu} \\
   \And
   Igor Markov \\
  Meta \\
   \texttt{imarkov@fb.com} \\
}

\begin{document}

\maketitle

\begin{abstract}
Today, Machine Learning (ML) applications can have access to tens of thousands of features. With such feature sets, efficiently browsing and curating subsets of most relevant features is a challenge. In this paper, we present a novel approach to visualize up to several thousands of features in a single image.
The image not only shows information on individual features, but also expresses feature interactions via the relative positioning of features.
\end{abstract}

\section{Introduction}

Features are fuel for Machine Learning. However, using too many features comes at a high computing cost and can even degrade performance due to overfitting. The task of feature selection, part of feature engineering, addresses this challenge. It is crucial in ensuring a high-quality dataset to train ML models.
However, humans are faced with decisions of which feature selection algorithm to use and what values of the algorithm's hyper-parameters. For example, they need to pick the number of features to keep or a stopping threshold for filter-based or forward-backward selection methods~\cite{CHANDRASHEKAR201416}. To support the developers in making such decisions, we need to have a user interface to browse or visualize the features to help them interpret the feature selection results and utilize their domain expertise to select which features to keep~\cite{krause2014infuse}.

\begin{wrapfigure}{R}{.43\textwidth}
  \vspace{-1em}
  \centering
  \includegraphics[width=.43\textwidth]{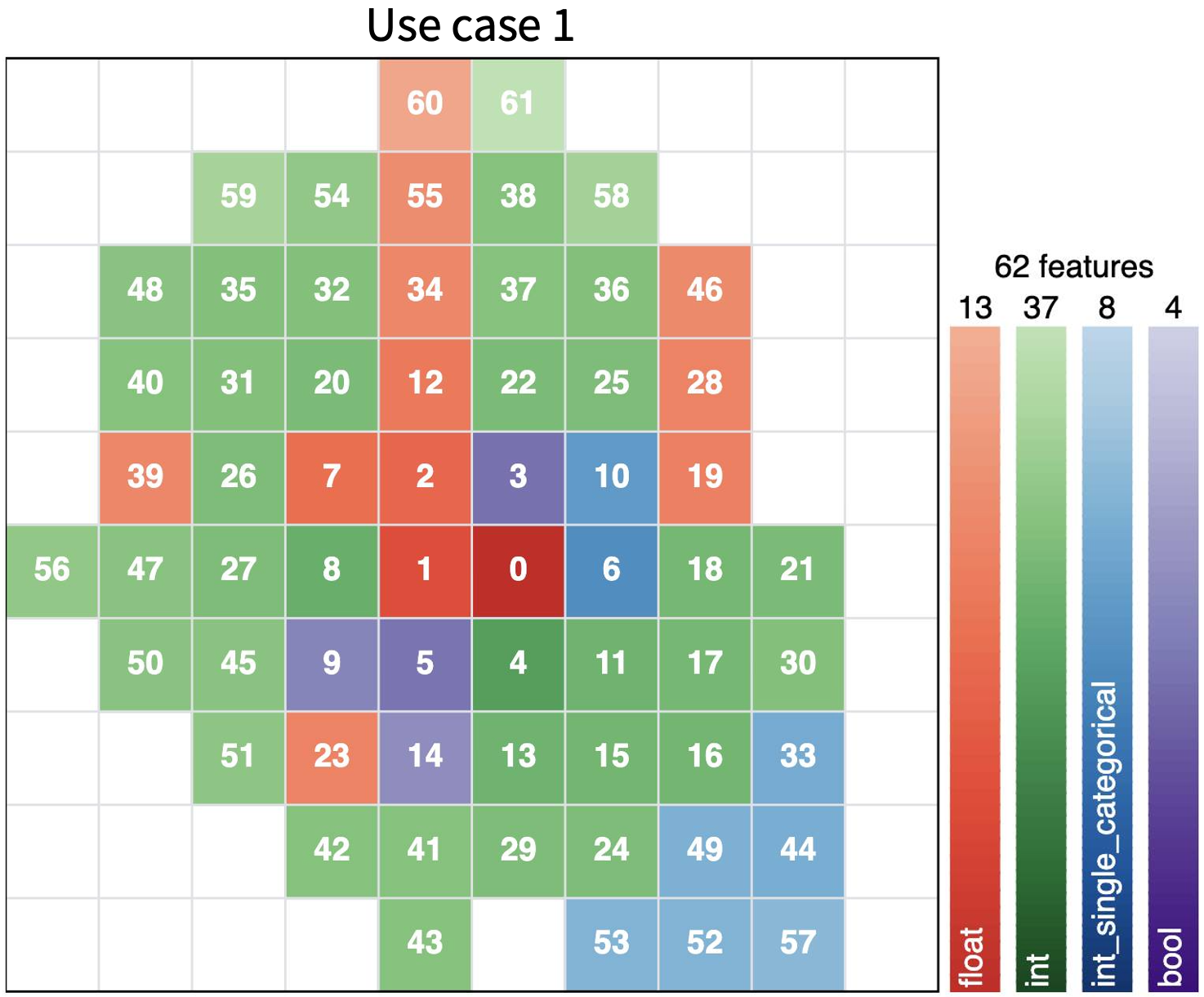}
  \caption{ML Features of a production use case visualized using \ourmethod. Each grid square represents a feature, with feature rank shown inside. The legend on the right gives colors of feature types and the number of features per type.}
  \label{fig:usecase1}
  \vspace{-2em}
\end{wrapfigure}

\begin{figure}[ht]
  \centering
  \includegraphics[width=\linewidth]{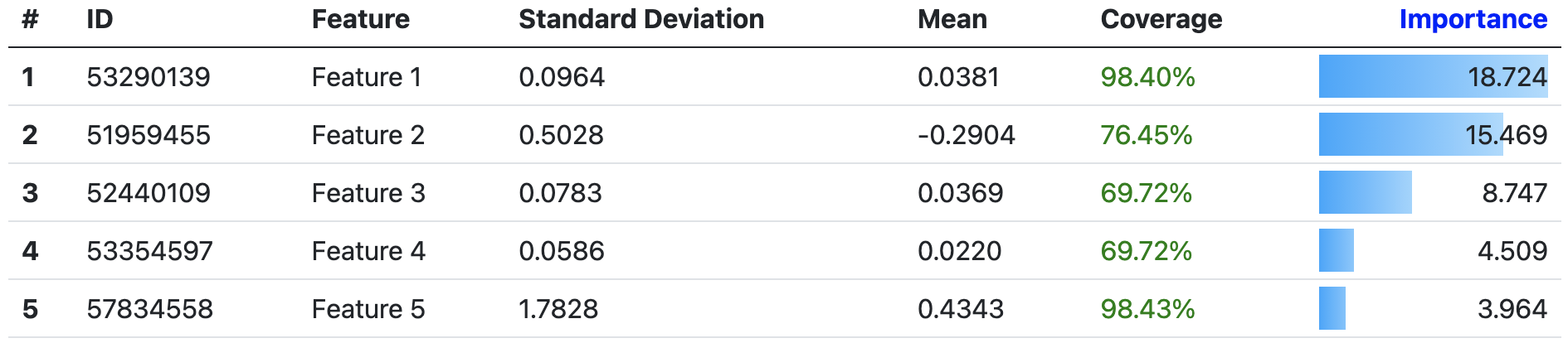}
  \caption{The beginning of a 21-page list with information on 420 features sorted by importance. This list is representative of how industry systems display such information. \hush{The name of features are replaced with mock-up data.}}
  \label{fig:feat_list}
  \vspace{-1em}
\end{figure}

Industry ML platforms typically display features in a list (Figure~\ref{fig:feat_list}) showing their importance scores along with statistics such as the mean, standard deviation, coverage. In practice, the list is long and spills into multiple pages, making it difficult to reason about the entire feature set. In addition, the importance scores exhibit the Long Tail phenomenon where the score quickly drops and the feature scores become quite similar. As a result, feature analysis becomes time-consuming and potentially inefficient.

We develop a novel method, called \ourmethod{}, to present ML features in a two-dimensional grid, in which each node represents a feature. Such visualization offers two main advantages: (1) efficiently displaying many features and their information in a single image compared to other representations such as lists or graphs, and (2) introducing an expressive spatial dimension capturing the interaction between features. Figure~\ref{fig:usecase1} shows ML features of a production use case visualized using \ourmethod{}.

\section{Our Approach}
\label{sec:approach}

We distinguish two important aspects in feature selection: (1) selecting individual features that are relevant/important to the problem, and (2) choosing sets of features that work together well. In particular, one should avoid redundancies and look for synergies between features. However, product engineers may find it difficult to reason about these aspects, and the current practice is to  trust automation tools, even though we have seen examples with relatively poor results.

To make feature information more accessible to a human, we develop a 2D visualization technique that represents each feature by a grid square (Figure \ref{fig:usecase1}), so that
\begin{itemize}
    \item color represents feature types,
    \item color saturation represents feature importance with respect to a given prediction task,\footnote{
    Raw feature importance values can be obtained from model-free methods~\cite{mrmr},
    GBDT models~\cite{ye2009stochastic}, or permutation-based metrics applied to arbitrary models.} normalized to the range [0, 255],
    \item numbers in grid squares rank features in the order they were selected.
\end{itemize}

Interactions between features
$f_i$ and $f_j$ are reflected in how features are assigned to grid squares. Specifically,  features that interact strongly 
tend to be placed close to each other. Pairwise interaction is expressed by a non-negative function $G(f_i , f_j)$, such as normalized co-occurrence of $f_i$ and $f_j$
% measured for an ML model 
used across different ML tasks in an ML platform~\cite{molino:declarative-ml} \hush{or used in the same decision tree in Random Forest model}
or, alternatively, Pearson's correlation of two features, which has shown surprising effectiveness during feature selection~\cite{mrmr}.
For a given function $G$, we determine grid positions $\{\pos{0}, \pos{1}, ... \}$ of features $\{f_0, f_1, ... \}$ (in the descending order of importance values) that minimize the following loss function consisting of the main term and two regularization terms:
\begin{equation}
%\begin{aligned}
    \label{eq:loss_fn}
    \mathcal{L} = \left( \sum_i I(f_i) \sum_{j:j<i} 
    G(f_i, f_j) \norm{\pos{i} - \pos{j}}_2^2 \right) + w_1 R_\text{center}  + w_2 R_\text{seq} ~~~~
     \text{subject to} \ \ \pos{i} \ne \pos{j} 
%\end{aligned}
\end{equation}
\vspace{-2mm}
\begin{equation*}
\text{where}~~~~~~~~~~~
R_\text{center} = \sum_{i} I(f_i) \norm{\pos{i}}_2^2 ~~~~~~~~~
R_\text{seq} = \sum_{i} I(f_i) \norm{\pos{i} - \pos{i-1}}_2^2
\end{equation*}

Here $I(f_i)$ is the importance score of feature $f_i$, whereas $w_1$ and $w_2$ are the weights of two regularization terms. The main term of $\mathcal{L}$ encourages nearby placement of features that interact strongly. However, in practice many pairs of features do not interact, which makes the optimization problem underdetermined. The two regularizers reduce the ambiguity, while making it easier to view and analyze features on the plot. The dominant term $R_\text{center}$ coerces important features toward the center, in particular, $p_0=[0, 0]^T$. The term $R_\text{seq}$ encourages features with similar importance ranks to be near each other, making them easier to find in the plot. The weights of the two regularizers are small ($w_1=0.05$ and $w_2=0.02$),
so that the main term drives the optimization.

Finding $\min \mathcal{L}$ is intractable yet unnecessary for visualization purposes.
Therefore, we greedily select $p_i$ ($p_i \neq p_j$) in the order of decreasing feature importance to minimize the proxy objective
\begin{equation}
%\begin{aligned}
    \label{eq:greedy_loss_fn}
    \mathcal{L}(i) = \left(\sum_{j: j < i} G(f_i, f_j) \norm{\pos{i} - \pos{j}}_2^2\right) + w_1 \norm{\pos{i}}_2^2 + w_2 \norm{\pos{i} - \pos{i-1}}_2^2\\
    % \text{s.t.} \ \ \pos{i} \ne \pos{j} \ \  \forall i \ne j 
%\end{aligned}
\end{equation}
This greedy strategy renders most important features in the center, consistently with $\min \mathcal{L}$, and allows us to drop $I(f_i)$ in the proxy objective.
Postprocessing $p_i$ with exhaustive-search (or branch-and-bound) optimization for small sets of nearby grid squares further improves results.

\section{Examples}

\begin{figure}[t]
% \subfloat[Use case 2 (420 features)\label{fig:usecase2}]{
%     \includegraphics[width=0.49\textwidth]{assets/usecase2}%
% }\hfill
% \subfloat[Use case 3 (1563 features)\label{fig:usecase3}]{
    \centering\includegraphics[width=0.49\linewidth]{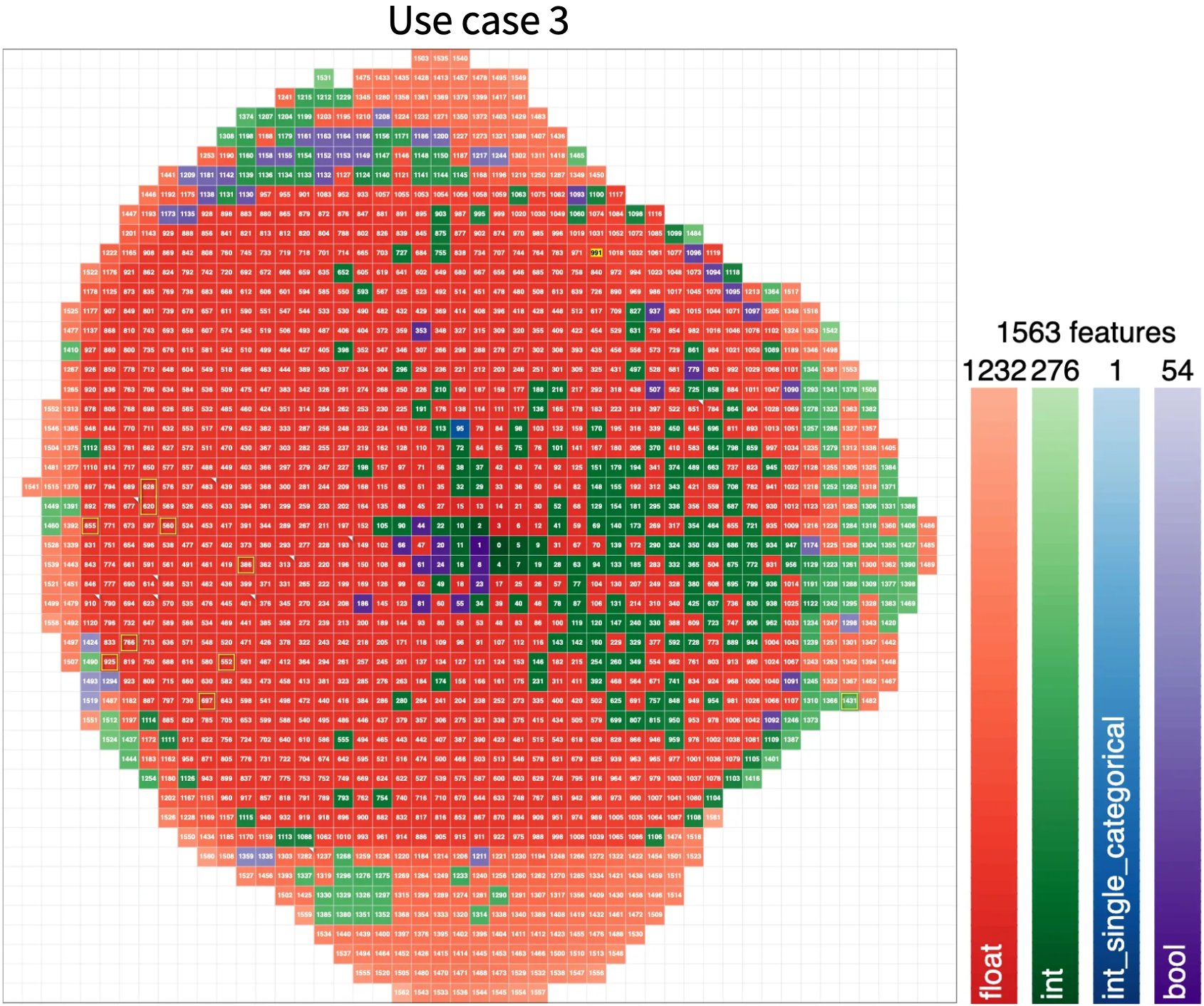}%
% }
% \caption{Features of two production use cases\hush{with numerous features} visualized with \ourmethod{}}
\caption{Features of a second production use case visualized with \ourmethod{}}
% \vspace{-1em}
\label{fig:usecase3}
\end{figure}

\begin{figure}[ht]
%\vspace{-1em}
\subfloat[
\label{fig:feat_highlight}
Two feature subsets: one selected by feature importance (yellow contour) and a handpicked one (white contour).
ML performance is shown on the upper right.
]{
    \includegraphics[width=0.48\textwidth]{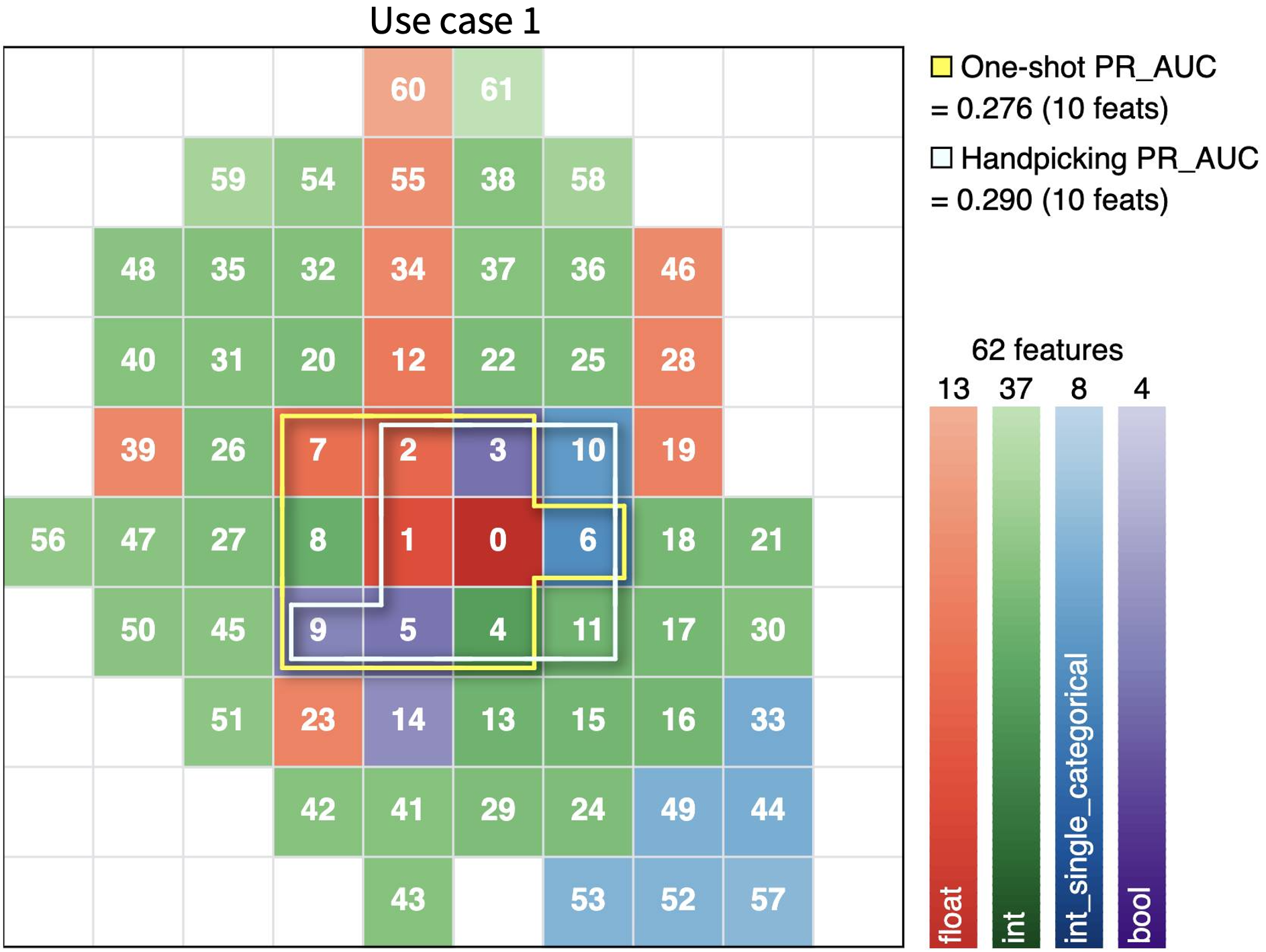}%
}\hfill
\subfloat[%\ourmethod{} visualizations are interactive.
Clicking on a feature pops up a window with additional information. The window closes when the mouse moves away. \label{fig:feat_popup}]{
    \includegraphics[width=0.48\textwidth]{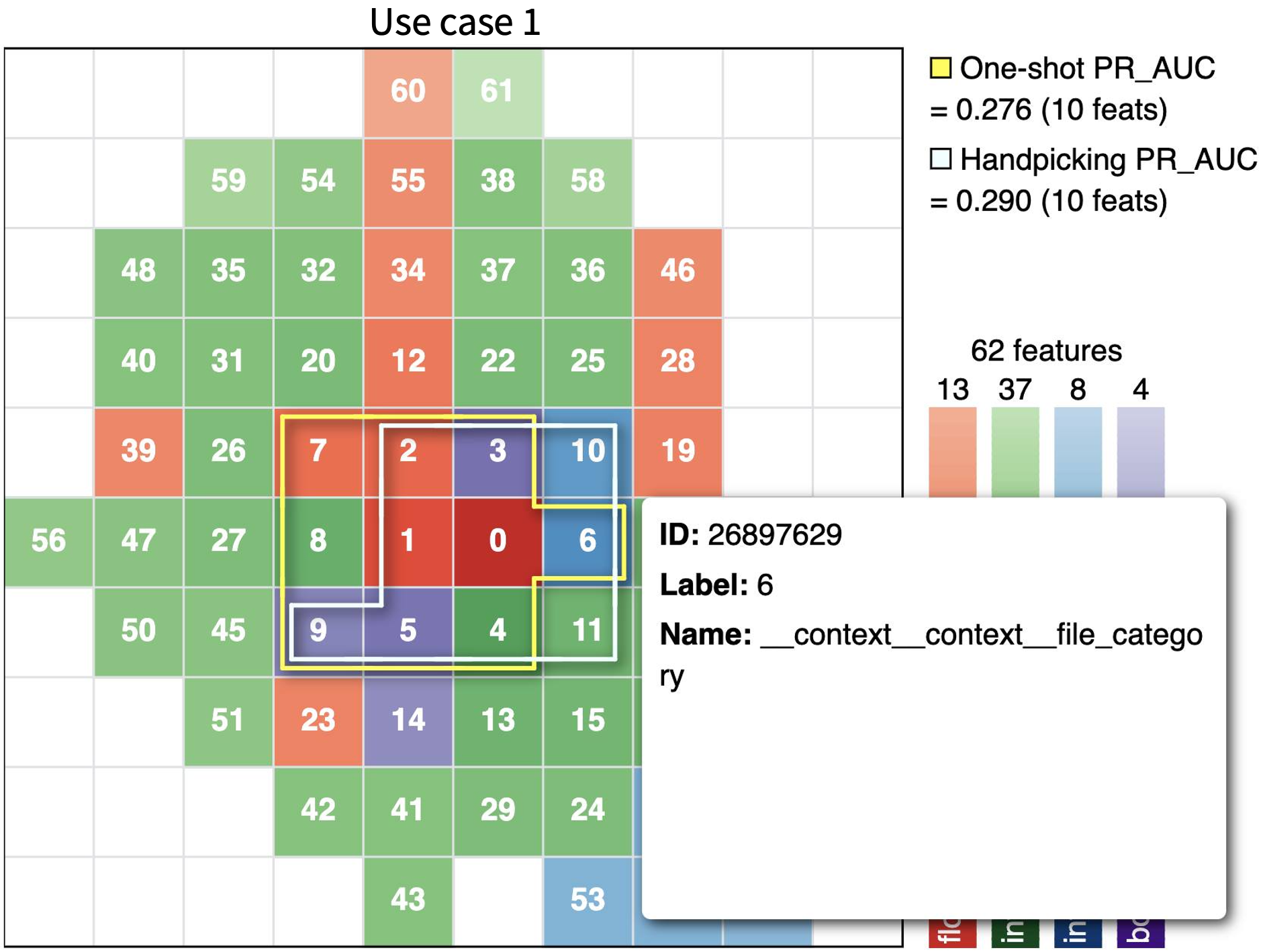}%
}
\caption{Interactive visualization with Picasso}
%\vspace{-1em}
\end{figure}

For illustrations 
 in Figures~\ref{fig:usecase1}\hush{,~\ref{fig:usecase2},} and~\ref{fig:usecase3}, we use two production use cases deployed on an ML platform with numbers of features from small to large. The results of different feature selection methods can be visualized side by side. Figure~\ref{fig:feat_highlight} shows the top 10 features automatically selected by feature importance (yellow contour) and 10 features selected manually (white contour) in one attempt based on the visualization of automatic selection. The features selected manually improve PR-AUC score by 1.37\%. When large feature subsets are disconnected in the plot, highlighting them with contours results in clutter, therefore \ourmethod{} can also mark individual grid squares with yellow dots. A simple heuristic to pick the method divides the area of the polygon containing the features (i.e., the number of features) by its perimeter. Given two large feature subsets, the one with the higher ratio is shown with contours, and the other one with dots.
In addition, clicking on a feature brings up additional information as illustrated in Figure~\ref{fig:feat_popup} for feature number 6. Our implementation using a Scalable Vector Graphics (SVG) backend will hide the pop-up box when the mouse moves away. % from the feature.

\section{Related Work}

Many visualization techniques have been developed to gain insights into data or ML models. In the context of data engineering, there are two main groups. 
% Methods in the first group visualize individual data points, such as t-SNE~\cite{van2008visualizing} that projects high-dimensional vectors into 2D or 3D continuous space. 
% The second visualizes features learned by a Neural Network model such as generating images that cause neurons in the model to fire strongly~\cite{DBLP:journals/corr/abs-1904-08939}.
Methods in the first group visualize individual data points or examples. For example, t-SNE~\cite{van2008visualizing} projects high-dimensional vectors into 2D or 3D continuous space; the Captum library~\cite{kokhlikyan2020captum} provides various algorithms to help interpreting a Neural Network model, e.g., finding and visualizing parts of input that the model uses to make prediction --- important pixels in an image or words in a document; and Activation Maximization methods~\cite{DBLP:journals/corr/abs-1904-08939} visualize features learned by a Neural Network model by generating images that cause neurons in the model to fire strongly.
The second group is more directly related to our work and includes visual analytics techniques that deal with entire features during feature selection~\cite{doi:10.1177/1473871620904671,Lu2017TheSI}. Compared to our work, they generally show a list of features along with their importance and statistical metrics~\cite{krause2014infuse,may2011guiding,muhlbacher2013partition}. They also allow users to filter, sort, plot the distribution of features~\cite{krause2014infuse} and perform pairwise feature analysis via plotting~\cite{muhlbacher2013partition} or computing correlation or mutual information~\cite{may2011guiding}.
However, as the number of features increases,
a simple list stretches beyond a single screen and is not particularly insightful. \ourmethod{} represents features as squares on a grid and displays numerous features compactly. Our visualizations leverage the geometric positions of features to express relationships between them such as correlation or other user-defined pairwise metrics of interaction~\cite{mrmr}.

As our technique plots features in 2D space, one
may draw some connections to feature embedding methods such as Feat2Vec~\cite{armona2019beyond}. However, work on feature embedding aims to find transformation functions that map the \textit{values} of features or examples from their original space into a new space. Hence, it is different from our method and is closer to methods in the first group such as t-SNE.
It may be able to be adapted to visualize a small/medium set of individual examples and may help the developers in designing their ML models. On the other hand, \ourmethod{} visualizes ML features and makes feature information more accessible to the developers to support doing feature selection. \ourmethod{} offers several unique capabilities (as per Section~\ref{sec:approach}) that are not available with other methods.

\section{Discussions and Future Work}

Our visualizations concisely display up to thousands of features while also expressing feature interactions. Several applications can benefit from this approach:
\begin{itemize}
    \item Visualizing features used in an ML model: as demonstrated in the examples, our tool can show hundreds of features compactly in a single image. It allows developers to quickly see the breakdown of features by type, feature importance, and their relationships. This may help developers gain more insights about the features used by their ML model.
    \item Helping ML engineers improve feature selection: our tool can visualize multiple results of feature selection on the same plot. Then, ML engineers can investigate why some features are chosen and some are not. Then, they can manually adjust the selected features to improve performance as illustrated in Figure~\ref{fig:feat_highlight}.
\end{itemize}

In our work, we have considered several alternatives to Equations \ref{eq:loss_fn} and \ref{eq:greedy_loss_fn}.
One such alternative is to {\em maximize} a sum of fractions
with distance terms in the denominator. This optimization tends
to be insensitive to the positions of the less important features and produces visualizations with less-than-reasonable positions on the fringes, e.g., we have seen examples with gaps between used grid cells. Another alternative is to replace the squared norms with plain norms. In practice, this makes little difference because (a) regularization terms only break ties left in the main term, (b) many $G$ terms are very close to zero and effectively discount large pairwise distances between such features, whereas features that interact tend to be placed close to each other.

Among further improvements, it is straightforward to highlight features from the same group,  e.g., embeddings of categorical features or features transformed from the existing features. Such information can help the developers to quickly see which transformations are useful and which aren't.
Additional filtering options can help the user select groups of features to plot or evaluate performance.
We also illustrate how interactivity helps to view additional feature information in pop-up windows, which can show many feature statistics.
% e.g., word embeddings, image pixels, etc. Such information can be valuable to developers. 
% We also illustrate how interactivity helps to view additional feature information; additional types of interactivity include zooming, viewing feature data/statistics, selecting groups of features, etc.

\newpage
\printbibliography

\end{document}